\documentclass[10pt,twoside,BCOR7mm,DIV15,headinclude,footexclude,cleardoubleempty,idxtotoc]{scrartcl}

\usepackage[english]{babel}
\usepackage{graphicx}
\usepackage{hyperref}
\usepackage{scrpage2}
\usepackage{hyperref}
\usepackage{ifthen}

\makeatletter
\renewcommand{\@biblabel}[1]{}
\renewcommand{\@cite}[2]{%
{#1\ifthenelse{\boolean{@tempswa}}{,#2}{}}}
\makeatother

\hypersetup{breaklinks=true
,colorlinks=true,linkcolor=black,urlcolor=blue
,citecolor=black}

\ofoot{\thepage}
\ifoot{}

\setheadsepline{1pt}

\setkomafont{pagehead}{\normalfont\sffamily}
\setkomafont{pagenumber}{\normalfont\rmfamily}

\usepackage{booktabs}
\usepackage{amsmath}
\usepackage{amssymb}
\usepackage{multicol}
\usepackage{float}

\makeatletter
\newcommand{\listofcontributions}{\@starttoc{con}}

\newcommand{\l@contribution} {\@dottedtocline{1}{1.5em}{2.3em}}
\makeatother

\newenvironment{contribution}{
\setcounter{section}{0}
\setcounter{figure}{0}
\setcounter{table}{0}
\begin{flushleft}
{\em Clumping in Hot Star Winds \\
W.-R.\ Hamann, A.\ Feldmeier \& L.\ Oskinova, eds.\\
Potsdam: Univ.-Verl., 2007 \\
URN: http://nbn-resolving.de/urn:nbn:de:kobv:517-opus-13981
} 
\end{flushleft}
}{
\newpage
\lehead{}
\rohead{}
}

\begin{document}

\setlength{\baselineskip}{2.5ex}

\begin{contribution}
% EXAMPLE AND TEMPLATE FILE FOR PROCEEDINGS OF THE CLUMPING WORKSHOP.
% PLEASE REPLACE THE TEMPLATE TEXT BY YOUR OWN ARTICLE.
% NOTE THAT YOU MUST NOT PROCESS THIS FILE, BUT THE MASTER FILE:
% latex masterfile; dvips masterfile

% RUNNING AUTHOR: PUT AUTHOR NAMED HERE
\lehead{R.\ Blomme}

% RUNNING TITLE; SHORTEN THE TITLE IF NECESSARY
% IN CASE OF A ONE-PAGE CONTRIBUTION (POSTER),
% SQUEEZE AUTHORS AND TITLE IN THIS LINE (Author: Title ...)
\rohead{Corotating Interaction Regions and clumping}

\begin{center}
% FULL TITLE HEADING
{\LARGE \bf Corotating Interaction Regions and clumping}\\
\medskip

% AUTHORS LIST
{\it\bf R. Blomme}\\

% AFFILIATIONS
{\it Royal Observatory of Belgium}\\

% ABSTRACT
\begin{abstract}
We present hydrodynamical models for Corotating Interaction Regions,
which were used by Lobel~(\cite{blomme:Lobel07}) to model the Discrete
Absorption Components in HD~64760. We also discuss our failure
to model the rotational modulations seen in the same star.
\end{abstract}
\end{center}

% TEXT OF THE PAPER, TWO-COLUMN STYLE
\begin{multicols}{2}

\section{Introduction}

Discrete Absorption Components (DACs) seen in ultraviolet lines 
are one of the clearest indicators of large-scale structure in the 
winds of hot stars.
Cranmer \& Owocki~(\cite{blomme:Cranmer+Owocki96}) 
attributed these DACs to
Corotating Interaction Regions (CIRs) which are formed when slow and
fast streams collide in the wind. In their model, CIRs
are produced by introducing a spot on the stellar surface.
The spot can literally be a spot, but can also simulate the effect
of non-radial pulsations, or a magnetic field.

In the present work, we attempt a {\em quantitative} fit to
the DACs seen in the Si {\sc iv} $\lambda$~1395
line of the B0.5 Ia star HD~64760. This 
paper describes the hydrodynamical models used for this fit and discusses
the problems in fitting the rotational modulations, which are an additional
feature seen in the P Cygni profiles of this star. In a companion
paper (Lobel~\cite{blomme:Lobel07}), the radiative transfer code
and the fitting of the observed DACs is discussed.

\section{Hydrodynamics}

To construct the hydrodynamical models, we use the Zeus3D code
(Stone \& Norman~\cite{blomme:Stone+Norman92}). 
We solve the 3-dimensional equations of hydrodynamics,
limited to the equatorial plane. This approximation 
is justified by the high $v$\,sin\,$i$ value of
HD~64760, which suggests that it is seen (nearly) equator-on.

In our model, we largely follow 
Cranmer \& Owocki~(\cite{blomme:Cranmer+Owocki96}).
We include the radiative acceleration due to the
spectral lines as well as the rotation of the star and the wind (through
angular momentum conservation). 
The energy equation is also solved (including radiative cooling and
a floor temperature), but this turned out to be unimportant in
the present calculations. We further introduce a circularly 
symmetric spot on the stellar surface, with its centre on the equator.
This spot is specified by its brightness $A_{\rm sp}$ and
its angular diameter $\Phi_{\rm sp}$. 

Contrary to Cranmer \& Owocki, we allow the spot to rotate with a
velocity $v_{\rm sp}$ {\em that can be different from the rotational
velocity} $v_{\rm rot}$. The main justification for this is that the
recurrence time of the DAC is measured to be 10.3~d while the rotation
period is only 4.1~d (see Lobel~\cite{blomme:Lobel07}). 
A similar assumption was also made
by Kaufer et al.~(\cite{blomme:Kaufer+al06}) 
in their model for the H$\alpha$ variations
in HD~64760.

\begin{figure}[H]
\begin{center}
\includegraphics[bb=30 20 537 528,clip,width=\columnwidth]{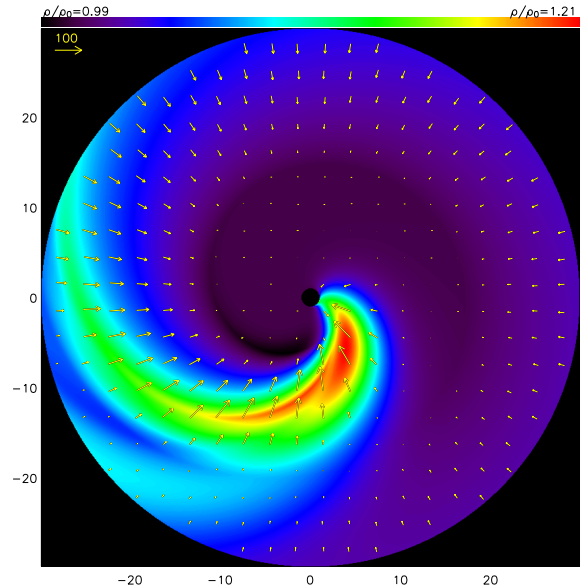}
\caption{The colour scale presents the density contrast of the CIR
with respect to the smooth wind. The velocity vectors plotted show
the differences with respect a smooth wind. The model parameters 
are: $A_{\rm sp}=0.1$, $\Phi_{\rm sp}=50^{o}$, 
$v_{\rm sp} = v_{\rm rot}/2.5$.
\label{blomme:fig CIR}}
\end{center}
\end{figure}

\begin{figure*}[!t]
\begin{center}
\includegraphics
  [width=0.9\textwidth]{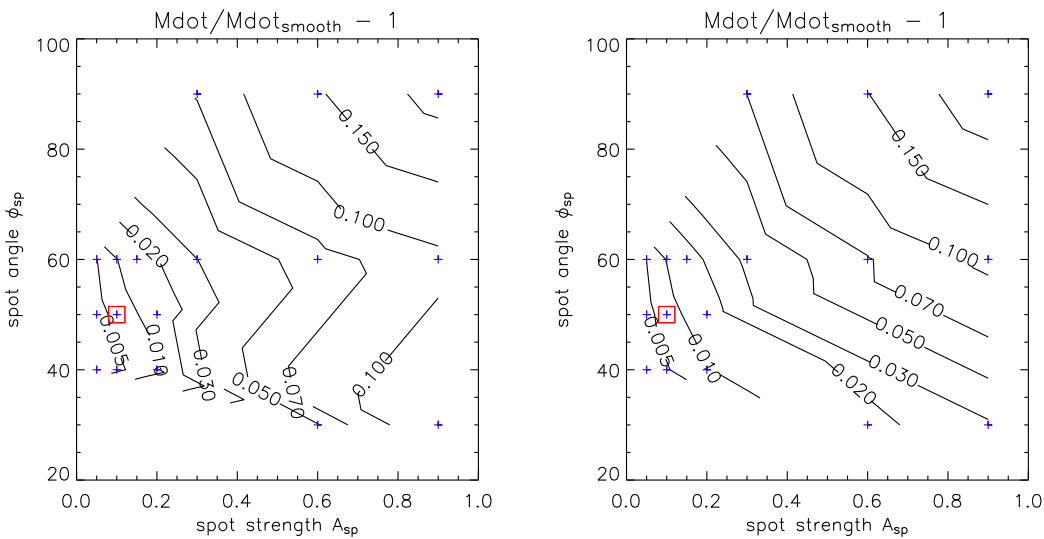}
\caption{Contour lines of the effect of a CIR on the mass-loss rate
as a function of spot brightness and angle. 
The plusses indicate the models we calculated. The best-fit model is
shown with a red square. The left panel shows the values measured
from the hydrodynamical models, the right panel the 
results from the approximation in Eq.~\ref{blomme:eq 2}.
\label{blomme:fig Mdot}}
\end{center}
\end{figure*}

The effect of a spot on the hydrodynamics of the wind
is easiest to understand if there is no rotation.
In that case, there is a sector above the spot with a
different wind (i.e. a
different mass-loss rate and terminal velocity) compared to the
smooth wind around it. When the star rotates, these different winds
collide and a spiral-shaped density enhancement is formed
(see Fig.~\ref{blomme:fig CIR}).
Trailing this density wave is a region with a
velocity plateau, which is 
mainly responsible for the formation of the DAC
(Lobel~\cite{blomme:Lobel07}). One should also note that the CIR
is not a stream of material particles, but is a pattern in the
wind. If we follow trace particles, we see that they can cross
the whole width of the CIR as they move out almost radially.

We constructed a large grid of hydrodynamical
models, from which the DACs were then
calculated using the {\sc Wind3D} code. The best fit to the
Si~{\sc iv} $\lambda$\,1395 line of HD~64760 was then determined
(see Lobel~\cite{blomme:Lobel07}). The hydrodynamical
model of this best fit is shown in Fig.~\ref{blomme:fig CIR}.

The total mass-loss rate of the structured wind
($\dot{M}_{\rm struct}$) is easy to calculate
if we assume the star to be non-rotating:
\begin{eqnarray}
\label{blomme:eq 1}
\dot{M}_{\rm struct} = \frac{\Omega}{4 \pi} \dot{M}_{\rm spot}
   + \left( 1 - \frac{\Omega}{4 \pi} \right) \dot{M}_{\rm smooth},
\end{eqnarray}
where $\Omega$ is the solid angle of the spot,
$\dot{M}_{\rm smooth}$ is the mass-loss rate in the smooth wind
and $\dot{M}_{\rm spot}$ that above the spot.
Using the approximate relation that $\dot{M} \propto L^{1/\alpha}$,
we can rewrite this as:
\begin{eqnarray}
\label{blomme:eq 2}
\frac{\dot{M}_{\rm struct}}{\dot{M}_{\rm smooth}} \approx
\frac{\Omega}{4 \pi} \left(1+A_{\rm sp} \right)^{1/\alpha}
   + \left( 1 - \frac{\Omega}{4 \pi} \right).
\end{eqnarray}
A rotating star will have a structure that is quite different from the
non-rotating model, but close to the stellar surface the differences
will be small. We can therefore use the density and velocity measured 
near the surface
to determine the mass-loss rate. The results for a series of models
are presented in Fig.~\ref{blomme:fig Mdot}. Our best-fit model for
the HD~64760 DAC corresponds to an additional mass loss of only 0.6~\%.

\section{Rotational modulations}

Encouraged by the good fit we obtained for the DAC, we also tried to fit
the ``rotational modulations" present in the dynamic spectrum. These are the
banana-shaped, nearly-flat absorption components seen in
Fig.~\ref{blomme:fig rot mods}. All our models fail however to reach the
terminal velocity sufficiently quickly. Our failure is surprising
in view of the good agreement presented by kinematical models such
as those by Owocki et al.~(\cite{blomme:Owocki+al95}) and 
Brown et al.~(\cite{blomme:Brown+al04}).

To investigate this further, we switched to using the kinematical
models developed by Brown et al. We find that these models
can indeed explain the
observations, provided the $v_{\rm rot}/v_\infty$ ratio is small
($\sim 0.05$). In such a case, the rotational modulation does reach
the terminal velocity sufficiently quickly (dotted white line in
Fig.~\ref{blomme:fig rot mods}). For HD~64760 however, we have that 
$v_{\rm rot}/v_\infty \approx 0.16$ and we obtain the solid white
line in Fig.~\ref{blomme:fig rot mods}, which does not fit the 
observations. Because we consider the spot velocity as a free parameter,
we can of course introduce a slower-moving spot (thereby getting
a more favourable $v_{\rm sp}/v_\infty$ ratio). However, the recurrence
time scale then becomes longer, stretching out the time-axis, and we
again obtain a bad fit which is indistinguishable from the
high $v_{\rm rot}/v_\infty$ solution (solid white line).

\begin{figure}[H]
\begin{center}
\includegraphics[bb=27  5 282 306,clip,width=\columnwidth]{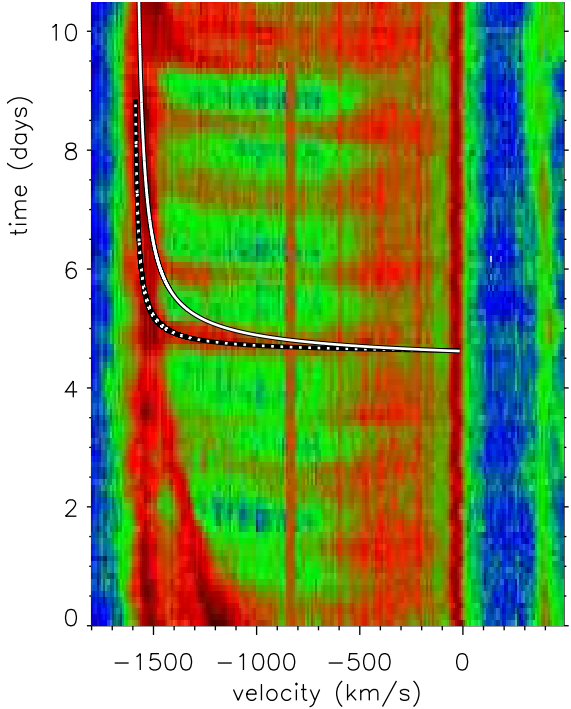}
\caption{The dynamic spectrum of the Si~{\sc iv} $\lambda$~1395 line
in HD~64760. One of the observed rotational modulations is fitted with
a kinematical model based on Brown et al.~(\cite{blomme:Brown+al04}).
The dashed white line has $v_{\rm rot}/v_\infty=0.05$, the solid one
has $v_{\rm rot}/v_\infty=0.16$ and is indistinguishable
from the $v_{\rm sp}/v_\infty=0.05$ solution.
\label{blomme:fig rot mods}}
\end{center}
\end{figure}

\section{Conclusions}

Lobel~(\cite{blomme:Lobel07}) and the present paper show that the DACs 
of HD~64760 can be well fitted with hydrodynamical models for CIRs.
The recurrence time of the 
HD~64760 DACs requires that the spots responsible for the CIRs
rotate considerably more slowly than
the stellar surface (by a factor of 2.5). The spots cannot be fixed on
the surface, thereby excluding magnetic fields as a possible cause.
As suggested by Kaufer et al.~(\cite{blomme:Kaufer+al06}),
a beat-pattern due to non-radial pulsations seems to be the only
valid explanation for the spot.

The additional mass-loss rate put into the wind is very small and
no shocks are formed due to the (weak) CIR in HD~64760. The CIRs
can therefore not be responsible for the clumping discussed at this
workshop, nor for the X-ray emission of these stars.
As Owocki~(\cite{blomme:Owocki98}) showed, CIRs are sensitive to the
amount of instability in the wind. Too large
an amount of clumping could therefore destroy the CIRs responsible for the DACs.
As the DACs are almost ubiquitous in hot-star winds, this imposes
important constraints on the amount of clumping.

Finally, our failure to reproduce the rotational modulations seen in 
HD~64760 suggests that we still lack a key component in our
understanding of these stellar winds.

{\bf Acknowledgements}  % Is this the correct font?
We thank Asif ud-Doula for making his version of the Zeus3D code available
to us.

% REFERENCES IN ALPHABETICAL ORDER

\end{multicols}

\end{contribution}

%%-------------------------------------------------------

\end{document}